\documentclass[aps,prl,amsmath,nofootinbib,twocolumn]{revtex4}
\usepackage[]{graphicx}
\usepackage{amsmath}
\usepackage{amsfonts}
\usepackage{amssymb}
\usepackage{bm}
\renewcommand{\mathbf}[1]{{\bm #1}}

\newcommand{\bn}{{\mathbf n}}
\newcommand{\bv}{{\mathbf v}}
\newcommand{\bnab}{{\mathbf \nabla}}

\newcommand{\be}{\begin{equation}}
\newcommand{\ee}{\end{equation}}
\newcommand{\ba}{\begin{eqnarray}}
\newcommand{\ea}{\end{eqnarray}}

\renewcommand{\section}[1]{\paragraph{\textbf{#1}}}

\begin{document}

\title{Anti-lensing: the bright side of voids}

\author{Krzysztof Bolejko$^1$, Chris Clarkson$^2$, Roy Maartens$^{3,4}$, David Bacon$^4$, Nikolai Meures$^4$, Emma Beynon$^4$ \\ ~
\\{\small\it $^1$Sydney Institute for Astronomy,
The University of Sydney, NSW 2006, Australia\\$^2$Centre for Astrophysics, Cosmology \& Gravitation,~and,~Department of Mathematics \& Applied Mathematics, 
University of Cape Town, Cape Town 7701, South Africa 
\\ $^3$Physics Department, University of the Western Cape, 
Cape Town 7535, South Africa
\\$^4$Institute of Cosmology \& Gravitation, 
University of Portsmouth, Portsmouth PO1 3FX, UK 
}}

\begin{abstract}
More than half of the volume of our Universe is occupied
by cosmic voids.
The lensing magnification effect from those under-dense regions is generally thought to give a small dimming contribution: objects on the far side of a void are supposed to be observed as slightly smaller than if the void were not there, which together with conservation of surface brightness implies net reduction in photons received. This is predicted by the usual weak lensing integral of the density contrast along the line of sight. We show that this standard effect is swamped at low redshifts by a relativistic Doppler term that is typically neglected. Contrary to the usual expectation, objects on the far side of a void are \emph{brighter} than they would be otherwise. Thus the local dynamics of matter in and near the void is crucial and is only captured by the full relativistic lensing convergence. There are also significant nonlinear corrections to the relativistic linear theory, which we show actually under-predicts the effect. We use exact solutions to estimate that these can be more than 20\% for deep voids. This remains an important source of systematic errors for weak lensing density reconstruction in galaxy surveys and for supernovae observations, and may be the cause of the reported extra scatter of field supernovae located on the edge of voids compared to those in clusters. 

\end{abstract}

\maketitle

\section{Introduction}

Lensing phenomena are measured not only
around virialized clusters of galaxies but also through and around unvirialized cosmic voids,
which occupy well above half the volume of the Universe. Here we show how the standard lensing magnification effect can be overwhelmed by relativistic corrections to the size and brightness of sources in and near voids.

\section{Magnification in the linear approximation}

The lensing magnification effect can be expressed in terms of the convergence
$\kappa$, which corrects the background angular diameter distance  ($\bar{d}_A$)
\be
d_A(z)=\bar d_A(z)[1-\kappa(z)]. \label{add}
\ee
The convergence in a perturbed 
$\Lambda$CDM universe is usually given as a line of sight integral over the density contrast $\delta$,
\begin{equation}
\kappa= \kappa_\delta =  \frac{3 }{2} H_0^2 \Omega_m \int_0^{\chi_S} {\rm d} \chi
\frac{ (\chi_S - \chi)}{\chi_S} \chi (1+z) \delta(\chi),
\label{WLF}
\end{equation}
where ${\rm d} \chi = {\rm d} z /H=-{\rm d}\eta$, $\chi$ is the comoving distance, $\eta$ conformal time and $S$ denotes the source.  In fact the full relativistic expression is~\cite{Bon2008,BDG2006} (see also \cite{extra})
 \be
\kappa= \kappa_{\nabla^2\Phi} +\kappa_v + \kappa_{SW} + \kappa_I,
\label{fgr}
 \ee
where the Sachs-Wolfe term $\kappa_{SW}$ is given by the difference in gravitational potential $\Phi$ between source and observer, and $\kappa_I$ is a line of sight integral over $\Phi$ and its conformal time derivatives $\Phi',\Phi''$. These two terms are sub-dominant \cite{Bon2008}, and we will not discuss their detailed form, although we do include them in our numerical calculations below. (The perturbed metric is 
${\rm d} s^2  = a^2\left[ -(1+2\Phi) {\rm d}\eta^2 +(1-2\Phi){\rm d}\mathbf{x}^2\right].$)

\begin{figure*}[ht!]
\begin{center}
\includegraphics[width=1\textwidth]{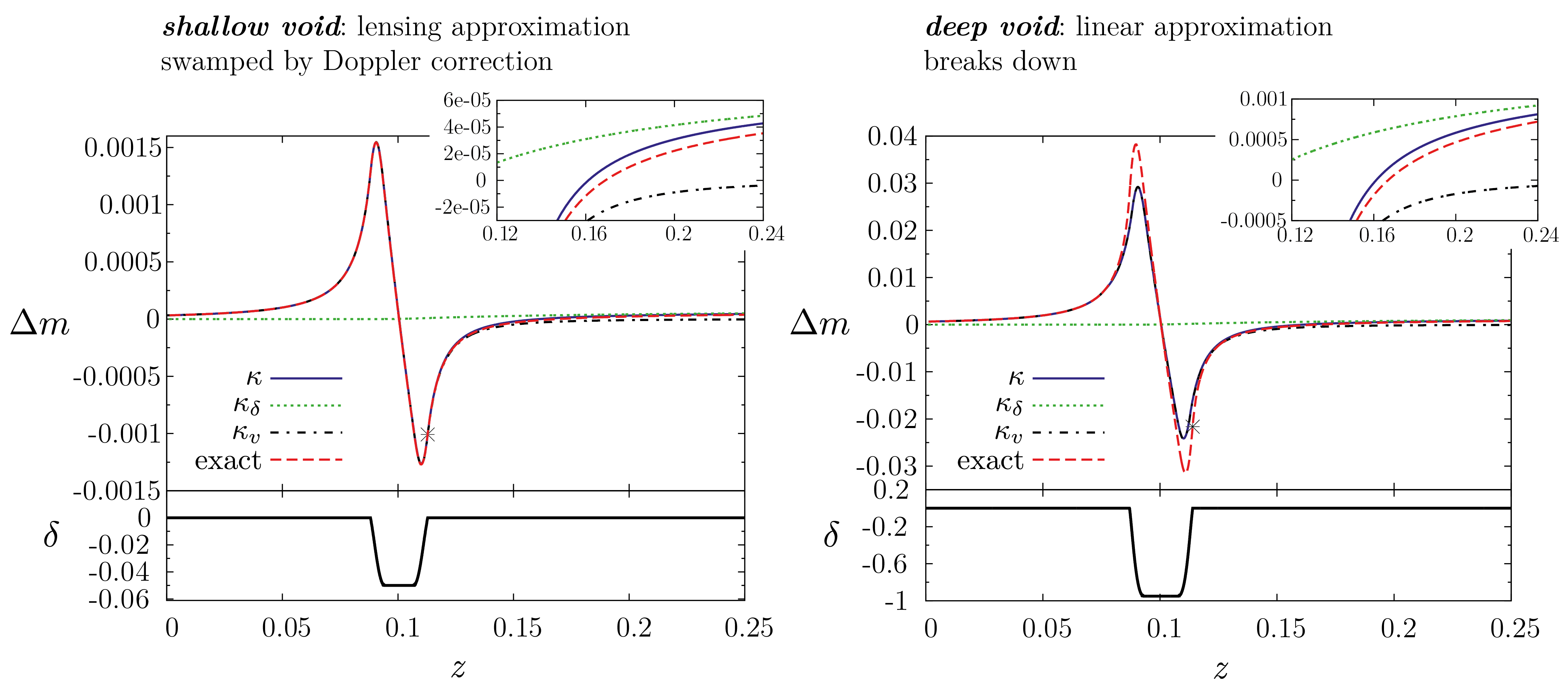}
\caption{ Change in distance modulus due to a 50 Mpc radius void, vs observed redshift, based on: relativistic convergence \eqref{fgr} (solid), usual weak lensing formula \eqref{WLF} (dotted),  Doppler term \eqref{VMF} (dot-dashed),
exact model (dashed) {\em (top panels)}.
There is a brightening for objects on the far side of the void, whereas the 
usual weak lensing predicts a dimming of much smaller magnitude (see insets). The left panel shows the effect for a small void well in the linear regime, and the right panel shows a very deep void where non-linear contributions increase the effect.  The bottom panels show the void density contrasts. (Asterisk marks the far edge of the void.) }
\label{fig1}
\end{center}
\end{figure*}

The usual form \eqref{WLF} is an approximation to the first term on the right of \eqref{fgr}, 
\begin{equation}
\kappa_{\nabla^2\Phi} = \int_0^{\chi_S} {\rm d} \chi
\frac{ (\chi_S - \chi)}{\chi_S} \chi  \nabla_\perp^2 \Phi.
\label{kphi}
\end{equation}
The screen-space Laplacian is $\nabla_\perp^2=\nabla^2-(\bn \cdot \bnab)^2-2\chi^{-1}\bn \cdot \bnab$, where $\bn$ is the unit direction from the source. The radial derivatives lead to terms proportional to $\Phi$, $\Phi'$ and $\Phi''$ \cite{BDG2006}, which are much smaller than the term $\nabla^2\Phi$ on the sub-Hubble scales of interest. Thus  in \eqref{kphi}, we may replace $\nabla_\perp^2\Phi$ by $\nabla^2\Phi$, which is given in terms of the density contrast $\delta$ by the Poisson equation. The general relativistic Poisson equation involves also the peculiar velocity ($v_i=\partial_iv$):
 \ba
\nabla^2\Phi &=& \frac{3 H_0^2 \Omega_m }{2a} \big( \delta-3aHv \big), \label{grp} \\
v &=& -\frac{2a}{ 3H_0^2\Omega_m}\big(\Phi'+aH\Phi \big). \label{oi}
 \ea
By \eqref{oi},  $aHv$ is of order $\Phi$ and may be neglected in \eqref{grp} on the relevant scales. Then \eqref{kphi} reduces to the usual lensing term \eqref{WLF}.
For an under-density, $\delta<0$, so that $\kappa_\delta<0$ if the underdensity is the dominant structure along the line of sight. Then \eqref{add} implies that the angular distance should be larger than the background value for a fixed $z$, and objects should consequently be observed to be smaller. Since surface brightness is  conserved in lensing, the total number of photons arriving from the object per unit time should be fewer. We can quantify this using the change to the distance modulus,
$\Delta m = 5 \log_{10} d_A/\bar{d}_A$, so that
a positive $\Delta m$ corresponds to a fainter source. We use $\Delta m$ rather than the convergence in the nonlinear examples below, since the relation between the two becomes complicated by the shear, which we do not consider explicitly here.

The usual formula is a good approximation to three of the terms in \eqref{fgr}:  $\kappa_\delta \approx \kappa_{\nabla^2\Phi} \gg \kappa_{SW},\kappa_I$. The remaining Doppler term, which arises from a shift in the redshift from its background value,
\be
\kappa_v=\left[1 - \frac{1+z_S}{\chi_S H_S }  \right] {\bv_S\cdot\bn},
\label{VMF}
\ee
is typically ignored -- but it cannot be neglected as emphasized by~\cite{Bon2008}, and as seen in Fig.~\ref{fig1}.

Figure~\ref{fig1} (left) shows the correction to the distance modulus for a spherical void of radius $50$ Mpc in the linear regime, $\delta_{\rm min}=-0.05$, located at $z=0.1$.  It is clear that $\kappa_\delta$ predicts a completely wrong magnitude for sources in or near the void -- and furthermore it predicts the wrong sign. By contrast, $\kappa_v$ gives a very good approximation to the full relativistic $\kappa$. Near the far edge of the void there is a significant \emph{positive} magnification signal: objects are brighter than they would be otherwise, an effect which extends far beyond the edge of the void. This is the opposite effect one expects based on a naive prediction using the usual lensing formula. { Note that $\kappa_v$ changes sign at the maximum of $\bar{d}_A(z)$, when the coefficient in \eqref{VMF} goes to zero: for higher redshifts ($z \gtrsim 1- 2$) the anti-lensing effect reverses and $\kappa_v$ reinforces $\kappa_\delta$.}

The right panel shows the prediction of an exact model of the void, using the Lemaitre-Tolman-Bondi (LTB) solution (see below). It demonstrates that the linear relativistic $\kappa$ is accurate for this amplitude of void. Since there is no background for the LTB case, the effect is not due to peculiar velocity, but rather to the extra redshifting of photons as they pass through a region of higher expansion rate and nonzero shear. The effect is stronger for a deep void with $\delta_\text{min}=-0.95$ (Fig.~\ref{fig1}, right panel), where the linear approximation underestimates the anti-lensing effect~-- we consider this in more detail below.

\section{Modelling voids via nonlinear solutions}

Real voids typically have $\delta\lesssim -0.8$ \cite{Sutter:2012wh}. { (This is for the galaxy density contrast: the typical underdensity in the total matter may be greater.)}
So we need to extend the relativistic perturbative analysis to deal with such voids. We can gain some insight via exact solutions of the Einstein field equations, where a void region is embedded in a homogeneous $\Lambda$CDM solution. We consider 3 models:

\noindent \underline{\it Spherical void}, using an LTB model, with the observer looking through the centre. The void can be compensated by a spherical shell of matter, or uncompensated. In the compensated case we choose
\begin{equation}
{\delta(r) \over \delta_\text{min}}  = \left\{ \begin{array}{ll}
1  &  r \leqslant \frac{1}{2} R, \\
-\frac{1}{2}r^{-2}R (r-R) {\rm e}^{3/2-6 [(r-R)/R]^2} &   \frac{1}{2} R \leqslant r \leqslant \frac{3}{2} R, \\
- r^{-2}\left( r-2R\right)^2  &  \frac{3}{2} R \leqslant r \leqslant 2 R, \\
0 &   r \geqslant {2} R, 
\end{array} \right.
\label{delta-com}
\end{equation}
\begin{figure}[h!]
\begin{center}
\includegraphics[width=1.0\columnwidth]{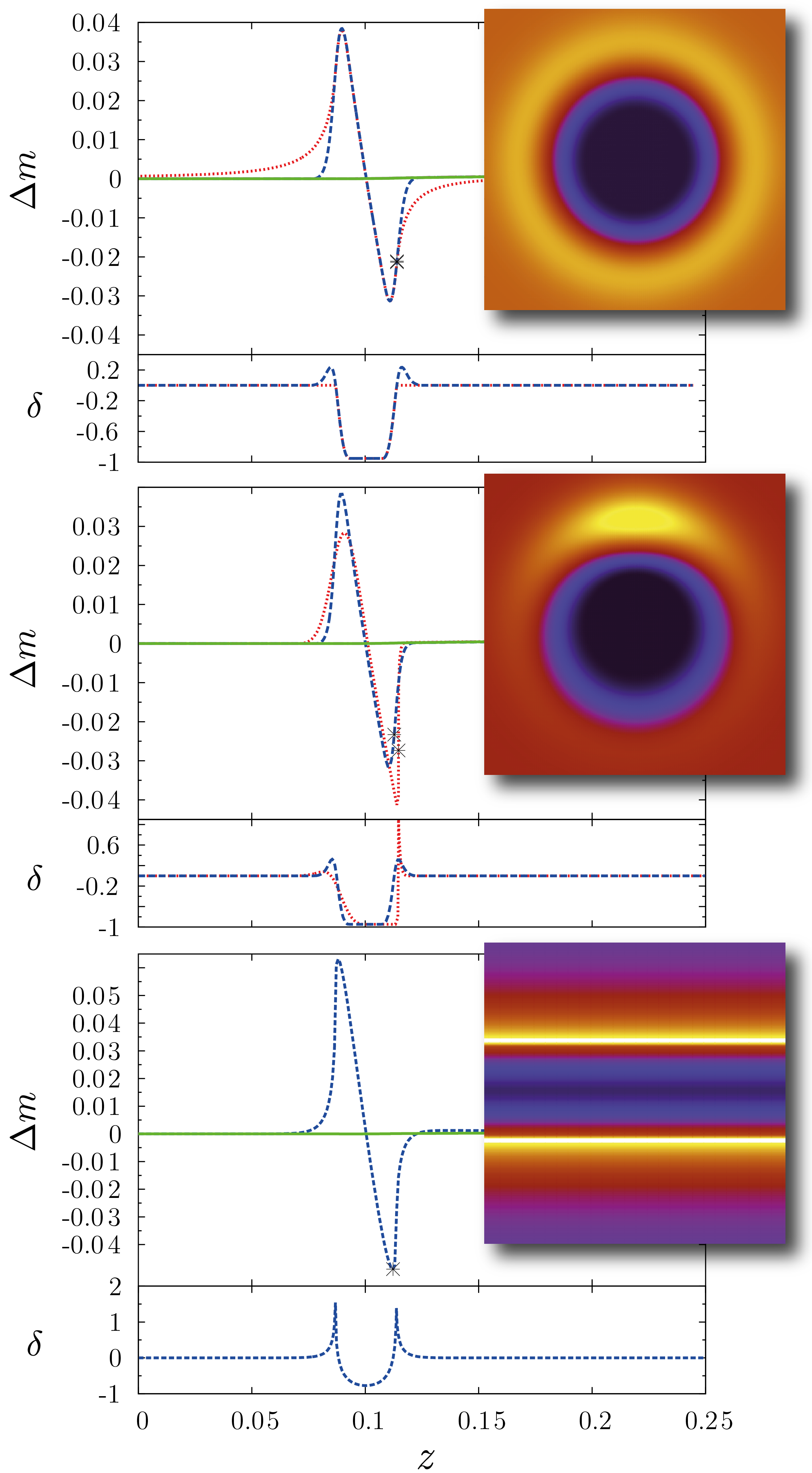}
\caption{Difference in magnitude (top panels) and density contrast (bottom panels) between the background $\Lambda$CDM model and the exact void models. We show results for an observer looking through 
spherical compensated (top), quasi-spherical (middle, shown for  both horizontal~-- blue, dashed~-- and vertical~-- red, dotted~-- lines of sight) and cylindrical (bottom) voids. The insets show the equatorial density contrast, with each square 160\,Mpc across. In the spherical case we also show the uncompensated case (red dotted line) discussed in Fig.~\ref{fig1}. The standard weak lensing prediction is shown in green in each case, which predicts the wrong sign and amplitude of the effect.
%
%
}
\label{fig2}
\end{center}
\end{figure}
and in the uncompensated case, 
\begin{equation}
{\delta(r) \over \delta_\text{min}}  = \left\{ \begin{array}{ll}
1  &  r \leqslant \frac{1}{2} R, \\
-\frac{1}{2}r^{-2}R (r-R) {\rm e}^{3/2-6 [(r-R)/R]^2} &   \frac{1}{2} R \leqslant r \leqslant  R, \\
0 &   r \geqslant  R. 
\end{array} \right.
\label{delta-unc}
\end{equation}
(We only consider growing modes, i.e. a uniform big bang time.)

\noindent  \underline{\it Quasi-spherical void}, with a mass concentration off to one side, with the observer looking either through the concentration, or along a line of sight not containing the concentration. We use the Szekeres type I model of \cite{SzT1a},
with the same mass distribution $M(r)$ as the compensated 
LTB model. In addition there is a dipole-like
contribution of the form $  - (S'/S)\cos \theta $  
where $S = r^{-0.99} {\rm e}^{-0.99 r/(2R)}$, for $r<2R$
and $S = (2R)^{-0.99} {\rm e}^{-0.99} = const$,  for $r \geq 2R$. 
This generates a compensated inhomogeneity extending to $r=2R$.

\noindent  \underline{\it Cylindrical void} with the observer looking across  the symmetry axis, using a Szekeres type II model  \cite{SzT2a}. The density profile orthogonal to the symmetry axis is compensated and extends to about $r=2R$. 

We standardize the voids in each model to have radius $50\,$Mpc, with centre at $z=0.1$. 
This corresponds to a slightly different comoving distance in each model, and the redshift of the near and far sides are slightly different. The depths we choose are $\delta_\text{min}=-0.95$ for the spherical and quasi-spherical cases, and $\delta_\text{min}=-0.8$ for the cylindrical case. Our model void size is typical in the Sloan Digital Sky Survey, where void radii are in the range 5 to 135$h^{-1}\,$Mpc \cite{Sutter:2012wh}.
Figure~\ref{fig2} shows the change in distance modulus for each type of void. In all cases we see the same qualitative behaviour as in the perturbative case: a relative dimming of objects on and near the closer side of the void, through to a relative brightening of objects located on and near the far side of the void. Details of the signal depend on the void shape and the nature of compensating regions, but 
the usual weak lensing prediction \eqref{WLF} is always completely wrong, unless the source is far from the void (cf the insets in Fig.~\ref{fig1}). It is interesting to note that for lines of sight which do not have overdensity that compensates explicitly for the void, the redshift range where the anti-lensing signal is significant is much larger. The reason is that the Hubble rate is larger than the background value beyond the region where $\delta$ returns to zero. This further illustrates the importance of modelling the dynamics of a void accurately to calculate the magnification correctly.

\begin{figure*}
\begin{center}
\includegraphics[width=1\textwidth]{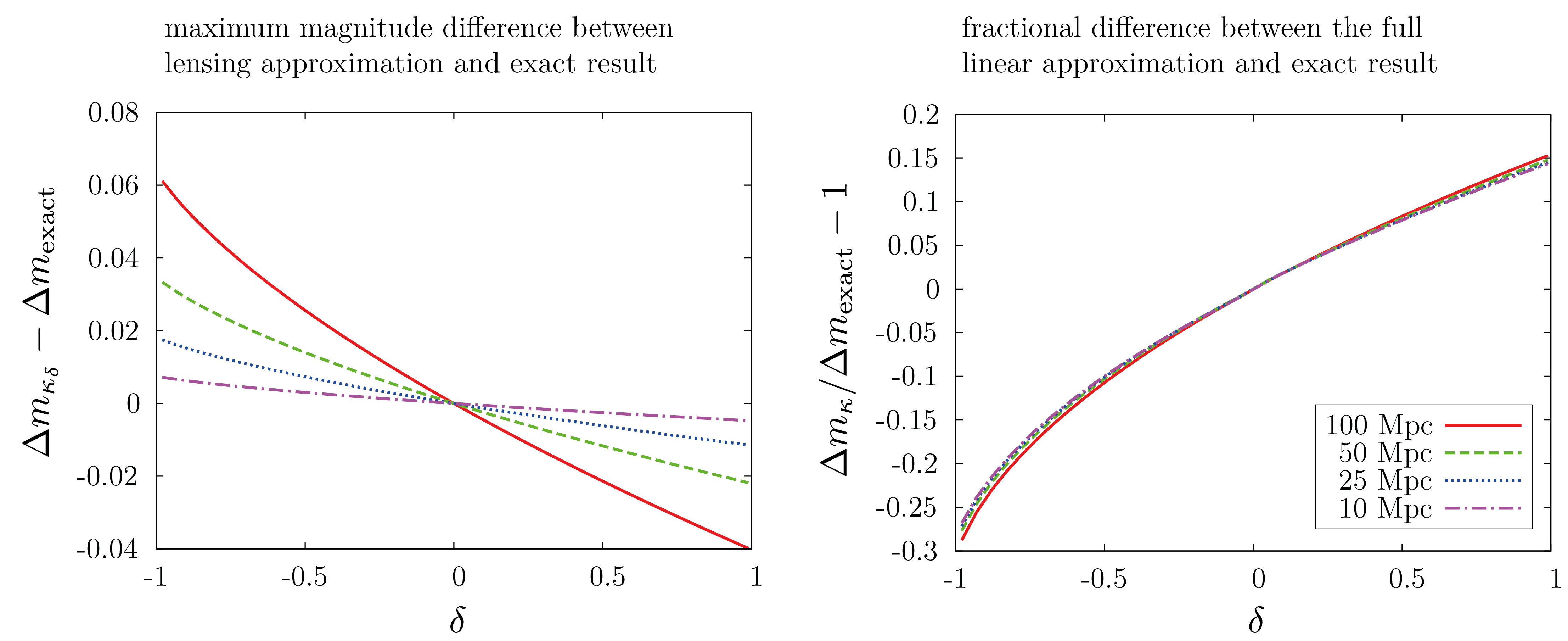}
\caption{(\emph{Left}) Accuracy of the lensing approximation compared to the exact LTB solution 
for the magnification 
near the far edge of a spherical compensated void ($\delta<0$) and a unvirialized over-dense lump ($\delta>0$). (\emph{Right}) Comparison of the full linear approximation to the exact result: maximum differences of $>20\%$ are seen for deep~-- but nevertheless realistic~-- voids.  Voids/lumps of radius 100, 50, 25, and 10 Mpc are used.}
\label{fig4}
\end{center}
\end{figure*}

Thus the predictions of the relativistic perturbative magnification persist in the nonlinear regime and are generic for different void configurations. 
But nonlinear effects can be large, and the linear relativistic analysis can be wrong by more than $\sim20\%$ compared to an exact spherical void model. This is shown in Fig. \ref{fig4}, which also includes the error for unvirialized overdensities.

\section{Discussion}

Our results illustrate a general principle: the measured magnitude of astronomical objects depends not only on internal properties of the source and statistics of large-scale structure, but also
on environment and the nature of inhomogeneity along the line of sight, which causes (de-)magnification. (See also \cite{Clarkson:2011br} 
and \cite{BF2012}.) We have used the full relativistic perturbative analysis to show that magnification of objects located in and near cosmic voids is dominated by the Doppler term \eqref{VMF} at low redshifts, which overwhelms the lensing effect \eqref{WLF} -- and is of opposite sign near the far edge of the void (see Fig.~\ref{fig1}). In other words, we have uncovered an anti-lensing effect for sources near the far side of voids. The usual weak lensing analysis fails completely to predict the foreground dimming and the background brightening from a void. Using exact solutions to model different voids, we have shown that this qualitative behaviour persists for nonlinear voids and for different shapes and ray directions (Fig.~\ref{fig2}). Nonlinear corrections to the relativistic linear predictions can be large, as shown in Fig.~\ref{fig4}. { In the left panel, note that the symmetry between negative and positive $\delta$ in the linear regime is broken as $|\delta |$ grows -- underdensity contrast is bounded below by -1, so that negative $\delta$ becomes more rapidly nonlinear than positive $\delta$.}

The failure of standard weak lensing for objects in and near under-dense void regions extends also to over-dense lump regions, provided that they are unvirialized. Common to both cases is the coherent flow into or out of the region, which sources a large velocity contribution. The key difference is that the lump occupies a much smaller (and shrinking) volume. As the lump continues to condense it tends to virialize and the correction then dies away, whereas the correction for the void {\em increases} with time.
For a source near a cluster of galaxies, there is no large net inflow or outflow since the structure is close to virialized, and so the usual weak lensing analysis is accurate. 

The relativistic linear analysis is accurate for computing the velocity contribution due to large-scale structure, provided the voids have $\delta_{\rm min}\gtrsim -0.2$ (see Fig.~\ref{fig4}). The velocity contribution from large-scale structure  was estimated for the angular power spectrum at fixed redshift in \cite{Bon2008} --  but without taking account of the nonlinear effects from voids that we have identified, which we have shown amplify the effect. The velocity contribution $C^v_\ell$ was predicted to exceed the usual $C^\delta_\ell$ for $z\lesssim 0.2$, to be about 50\% for $z\approx 0.5$ and to be negligible for $z>1$. The nonlinear corrections that we have identified due to voids with $\delta_{\rm min}\lesssim -0.2$ will introduce a systematic error into the perturbative calculation. A key question is: how to estimate the nonlinear void correction and thus correct for this systematic in galaxy surveys? A similar question applies also to supernovae magnitudes \cite{Clarkson:2011br}.

While the void effect on convergence is large and of the opposite sign to what is expected from lensing, the same is not true for lensing shear. The Doppler effect will move the redshift of a sheared source, which is a correction that should be taken into account in making density maps from shear data. However, the sign of the shear is unchanged, as the Doppler term does not distort an object's shape, but only changes its inferred size. As pointed out in \cite{Bon2008}, measurement of the convergence and lensing shear will provide a powerful probe of peculiar velocities.

Could this effect have already been detected? It is reported in~\cite{Sullivan:2010mg} that the scatter in the Hubble diagram of  SNIa magnitudes depends on the environment of the host galaxy. They find that galaxies which have a lower star formation rate are associated with SNIa which have a smaller scatter in the Hubble diagram (at $2-3\sigma$). As lower star formation rates are associated with galaxies in clusters compared to those in the field~-- the latter being more likely to be on the edge of a void~-- it may be that this extra scatter is due to the large relativistic anti-lensing effect we have described here. This deserves further investigation.

%

~\\{\bf Acknowledgments:}\\
We thank Mat Smith for discussions.
DB and RM are supported by the UK Science \& Technology Facilities Council (grant no. ST/H002774/1). DB, CC, RM, NM are supported by a Royal
Society (UK)/ National Research Foundation (SA) exchange grant. RM is supported by the South African Square Kilometre
Array Project. CC and RM are supported by the National Research Foundation.

\end{document}